\documentclass[3p,twocolumn,preprint]{elsarticle}
\pdfoutput=1
\usepackage[utf8]{inputenc}
\usepackage[english]{babel}
\usepackage[T1]{fontenc}
\usepackage{amsmath}
\usepackage{hyperref}
\usepackage{csquotes}
\usepackage{tikz}
\usepackage{lipsum}
\usepackage{times,amsmath,amsfonts,amssymb,latexsym}
\usepackage{graphicx,epsf}
\usepackage{subfigure}
\biboptions{sort&compress}
\newcommand{\be}{\begin{equation}}
\newcommand{\ee}{\end{equation}}
\newcommand{\ba}{\begin{eqnarray}}
\newcommand{\ea}{\end{eqnarray}}
\newcommand\tr{{\operatorname{tr}}}
\newcommand{\ignore}[1]{}
\newcommand{\otoc}[0]{\text{OTOC}}

\newcommand{\ket}[1]{\left | {#1} \right \rangle }

\newcommand{\bra}[1]{\left \langle {#1} \right | }

\newcommand{\de}[0]{{\operatorname{d}}}

\newcommand{\aver}[1]{ \left\langle  {#1}  \right\rangle }

\newtheorem{prop}{Proposition}
\def\norm#1{\Vert #1\Vert}
\def\CC{{\rm\kern.24em \vrule width.04em height1.46ex depth-.07ex
    \kern-.29em C}}
\def\P{{\rm I\kern-.25em P}}
\def\RR{{\rm
         \vrule width.04em height1.58ex depth-.0ex
         \kern-.04em R}}
\def\bbbone{{\mathchoice {\rm 1\mskip-4mu l} {\rm 1\mskip-4mu l}
{\rm 1\mskip-4.5mu l} {\rm 1\mskip-5mu l}}}
\def\bbbc{{\mathchoice {\setbox0=\hbox{$\displaystyle\rm C$}\hbox{\hbox
to0pt{\kern0.4\wd0\vrule height0.9\ht0\hss}\box0}}
{\setbox0=\hbox{$\textstyle\rm C$}\hbox{\hbox
to0pt{\kern0.4\wd0\vrule height0.9\ht0\hss}\box0}}
{\setbox0=\hbox{$\scriptstyle\rm C$}\hbox{\hbox
to0pt{\kern0.4\wd0\vrule height0.9\ht0\hss}\box0}}
{\setbox0=\hbox{$\scriptscriptstyle\rm C$}\hbox{\hbox
to0pt{\kern0.4\wd0\vrule height0.9\ht0\hss}\box0}}}}
\def\bbbz{{\mathchoice {\hbox{$\sf\textstyle Z\kern-0.4em Z$}}
{\hbox{$\sf\textstyle Z\kern-0.4em Z$}}
{\hbox{$\sf\scriptstyle Z\kern-0.3em Z$}}
{\hbox{$\sf\scriptscriptstyle Z\kern-0.2em Z$}}}}

\newtheorem{theorem}{Theorem}

\begin{document}
\begin{frontmatter}
\title{Isospectral twirling and quantum chaos}
\author[1]{Lorenzo Leone\corref{cor1}}
\ead{Lorenzo.Leone001@umb.edu}
\cortext[cor1]{Corresponding author}
\author[1]{Salvatore F.E. Oliviero}
\author[1]{Alioscia Hamma}
\address[1]{Physics Department,  University of Massachusetts Boston,  02125, USA}
\begin{abstract}
We show that the most important measures of quantum chaos like frame potentials, scrambling, Loschmidt echo and out-of-time-order correlators (OTOCs) can be described by the unified framework of the isospectral twirling, namely the Haar average of a $k$-fold unitary channel. We show that such measures can then be always cast in the form of an expectation value of the isospectral twirling. In literature, quantum chaos is investigated sometimes through the spectrum and some other times through the eigenvectors  of the Hamiltonian generating the dynamics. We show that,
by exploiting random matrix theory, these measures of quantum chaos
clearly distinguish the finite time profiles  of probes to quantum chaos corresponding to chaotic spectra given by the Gaussian Unitary Ensemble (GUE) from the integrable spectra given by Poisson distribution and the Gaussian Diagonal Ensemble (GDE). On the other hand, we show that the asymptotic values do depend on the eigenvectors of the Hamiltonian.  We see that, the isospectral twirling of Hamiltonians with  eigenvectors  stabilizer states, does not possess chaotic features, unlike those Hamiltonians whose eigenvectors are taken from the Haar measure. As an example, OTOCs obtained with Clifford resources decay to higher values compared with universal resources. Finally, we show a crossover in the OTOC behavior between a class of integrable models and quantum chaos.
\end{abstract}
\end{frontmatter}
%%%%%%%%%%Abstract

\section{Introduction} 
The onset of chaotic dynamics is at the center of many important phenomena  in quantum many-body systems. From thermalization in a closed system\cite{lloyd1988black,rigol2008thermalization, santos2010onset,popescu2006entanglement,srednicki1994chaos,reimann2007typicality,eisert2015quantum,polkovnikov2011colloquium,garcia2018relaxation,tasaki2016typicality,reimann2015generalization} to scrambling of information in quantum channels\cite{hosur2016chaos,ding2016conditional,brown2012scrambling,liu2018entanglement,liu2018generalized,styliaris2020information}, black hole dynamics\cite{hayden2007black,shenker2015stringy,kitaev2014hidden,cotler2017black}, entanglement complexity\cite{yang2017entanglement,chamon2014emergent}, to pseudorandomness in quantum circuits\cite{harrow2009random,gharibyan2018onset,brown2010random,brown2010convergence,nahum2018operator} and finally the complexity of quantum evolutions\cite{brown2018second,dowling2008geometry,zhou2017operator,benenti2009complex}. Several probes of quantum chaos have been studied in recent years \cite{maldacena2016bound, lashkari2013towards,xu2019locality,anand2020quantum}. 
 Chaos, equilibration, thermalization and other related phenomena are described by the behavior of entanglement growth and typicality, Loschmidt Echo, and to out-of-time-order correlation functions (OTOCs)\cite{larkin1969quasiclassical,kitaev2014hidden,lin2018out,chenu2019work,touil2020quantum,happola2012universality,swingle2016measuring,von2018operator,swingle2018unscrambling}. Information scrambling is characterized by the tripartite mutual information\cite{hosur2016chaos,ding2016conditional} and its connection OTOCs. Pseudorandomness is characterized by the frame potential which describes the adherence to moments of the Haar measure\cite{scott2008optimizing,roberts2017chaos}; the complexity of entanglement is characterized by the adherence to the random matrix theory distribution of the gaps in the entanglement spectrum\cite{yang2017entanglement,chamon2014emergent}.

%time evolution by Gaussian Unitary Ensemble (GUE) Hamiltonians and
%analytically compute out-of-time-ordered correlation functions (OTOCs) and frame
%potentials to quantify scrambling, Haar-randomness, and circuit complexity

Random matrix theory (RMT) has been extensively studied and applied to quantum chaos\cite{wigner1951statistical,haake1991quantum,mehta2004random, tao2012topics, rao2020wigner,chen2018operator}. The quantization of classical chaotic systems has often resulted in quantum Hamiltonians with the same level spacing statistics of a random matrix taken from the Gaussian Unitary Ensemble (GUE). One could take the behavior of OTOCs, entanglement, frame potentials and other probes under a time evolution induced by a chaotic Hamiltonian, that is, e.g. a random Hamiltonian from GUE and define it as the characteristic behavior of these quantities for quantum chaos\cite{cotler2017chaos, hunter2018chaos,Balasubramanian_2014}. Though we agree with the heuristics of this approach, it would be important to compare the time behavior of these probes in systems that are not characterized by a spectrum given by random matrix, or, on the other hand by Hamiltonians whose eigenvectors are not a random basis according to the Haar measure, e.g., Hamiltonians with  eigenvectors that, although possessing high entanglement, do not contain any magic, that is, they are stabilizer states. Attempts at showing the difference in behavior between chaotic and non-chaotic behavior are often limited to specific examples \cite{bao2020out,lin2018out}. Moreover, given the proliferation of probes to quantum chaos, one does feel the necessity of having a unified framework to gather together all these results. 

In this paper, we set out to provide such a unifying framework and to prove that one can clearly distinguish chaotic from non-chaotic dynamics. The framework is provided by the {\em isospectral twirling} $  \hat{\mathcal{R}}^{(2k)}(U)$, that is, the Haar average of a $k-$fold channel. This operation randomizes over the eigenstates of a unitary channel $U$ but leaves the spectrum invariant. In this way, one obtains quantities that are functions of the spectrum only. The unitary channel represents the quantum evolution induced by a Hamiltonian. Chaotic Hamiltonians feature spectra obeying the random matrix theory, e.g. GUE, while integrable systems possess spectra obeying other statistics \cite{scaramazza2016integrable, riser2020power, prakash2020universal,riser2020nonperturbative}. The main results of this paper are: (i) the isospectral twirling unifies all the fundamental probes $\mathcal P$ used to describe quantum chaos in the form of  $\aver{\mathcal P_{\mathcal O}}_{G} =\tr[ \tilde T \mathcal O \hat{\mathcal{R}}^{(2k)} ]$, where $\tilde{T}$ is a rescaled permutation operator, $\mathcal O$ characterizes the probe, $\aver{\cdot}_{G}$ is the Haar average and (ii) by considering the isospectral twirling associated to a $k-$doped Clifford group $\mathcal C(d)$, we show that the asymptotic temporal behavior of the OTOCs interpolates between a class of itegrable models and quantum chaos, and does not depend on the specific spectrum of the Hamiltonian; (iii) finally, by computing the isospectral twirling for the spectra corresponding to the chaotic Hamiltonians in GUE and integrable ones - Poisson, Gaussian Diagonal Ensemble (GDE) - the isospectral twirling can distinguish chaotic from non chaotic behavior in the temporal profile of the probes, though all the spectra lead to the same asymptotic behavior - a sign of the fact that chaos is not solely determined by the spectrum of the Hamiltonian, but also by its eigenvectors.
\section{Isospectral Twirling}
Let $\mathcal{H}\simeq\mathbb C^d$ be a $d-$dimensional Hilbert space and let $U\in \mathcal{U}(\mathcal{H})$ with spectral resolution $U=\sum_{k}e^{-iE_k t}\Pi_k$, where $\Pi_{k}$ are orthogonal projectors on $\mathcal{H}$. We can think of the  $\text{Sp}(H)\equiv\{E_{k}\}_{k=1}^{d}$ as the spectrum of a Hamiltonian $H$. 
Through $H$ we can generate an isospectral family of unitaries $\mathcal{E}_H\equiv \{U_G (H)\}_G:=\{ G^\dagger \exp\{-iHt\} G,\,G\in\mathcal{U}(\mathcal{H})\}$.
Define {\em isospectral twirling} the $2k-$fold Haar channel of the operator $U^{\otimes k,k}\equiv U^{\otimes k}\otimes U^{\dag\otimes k}$ by
\ba
  \hat{\mathcal{R}}^{(2k)}(U):=\int\, dG\, G^{\dag \otimes 2k}\left(U^{\otimes k,k}\right)G^{\otimes 2k}
\label{isospectraltwirling}
\ea
where $dG$ represents the Haar measure over $\mathcal{U}(\mathcal{H})$. This object has been previously used to demonstrate convergence to equilibrium under a random Hamiltonian\cite{brandao2012convergence} or the behavior of random quantum batteries\cite{caravelli2020random}. Under the action of (\ref{isospectraltwirling}), the spectrum of $U$ is preserved.
A general way to compute  the above average is to use the Weingarten functions\cite{collins2003moments}. We obtain:
\ba
\hat{\mathcal{R}}^{(2k)}(U)=\sum_{\pi\sigma} (\tilde{\Omega}^{-1})_{\pi\sigma}\tr(\tilde{T}_{\pi}^{(2k)}U^{\otimes k,k})\tilde{T}_{\sigma}^{(2k)}
\ea
where $\tilde{T}_{\pi}^{(2k)}\equiv T_{\pi}^{(2k)}/d_{\pi}^{(2k)}$,  $\pi,\sigma\in S_{2k}$ are (rescaled) permutation operators of order $2k$, $d_{\pi}^{(2k)}=\tr T_{\pi}^{(2k)}$ and $(\tilde{\Omega}^{-1})_{\pi\sigma}\equiv [\tr(\tilde{T}_{\pi}^{(2k)} \tilde{T}_{\sigma}^{(2k)})]^{-1}$ are rescaled Weingarten functions. Notice that, through $U$, the isospectral twirling is a function of the time $t$. 

Taking the trace of Eq.\eqref{isospectraltwirling} one obtains the $2k-$point spectral form factors: 
 $
\tr(\hat{\mathcal{R}}^{(2k)}(U))=|\tr(U)|^{2k} =(d+Q(t))^{k}
$
which follows easily from the cyclic property of the trace and the fact that $\int dG=1$. The object $Q(t)=\sum_{i\neq j}\cos[(E_i-E_j)t]$\cite{caravelli2020random} is related to the quantum advantage of the performance of random quantum batteries.
For $k=1,2$ these spectral form factors read  $|\tr(U)|^{2} =\sum_{i,j}e^{i(E_i-E_j)t}$ and $|\tr(U)|^{4} =\sum_{i,j,k,l}e^{i(E_i+E_j-E_k-E_l)t}$. More generally, consider the coefficients $\tilde{c}^{(2k)}_\pi (U) :=\tr(\tilde{T}_{\pi}^{(2k)}U^{\otimes k,k})$. After the twirling, all the information about the spectrum of the Hamiltonian $H$ is encoded in the $\tilde{c}_{\pi}^{(2k)}(U)$. We see that the $2k-$point spectral form factors come from the identity permutation $\pi =e$ such that $T_{e}^{(2k)} = \bbbone^{\otimes 2k}$. For $k=2$ and the permutation $T_{\pi}^{(4)} = T_{(12)(3)(4)}^{(4)}\equiv T_{(12)}^{(4)}$ we instead obtain another spectral form factor, namely  $\tilde{c}^{(4)}_{(12)} (U) =d^{-3}\tr(U^2)\tr(U^\dag)^2 =d^{-3}\sum_{i,j,k}e^{i(2E_i-E_j-E_k)t}$, which we will be needing later. Spectral form factors only depend on the spectrum of $U$. In particular, those we listed only depend on the gaps in the spectrum of $H$. For $k=2$, we set up this lighter notation for objects that we will be using a lot: $\tilde{c}^{(2)}_e\equiv\tilde{c}_2$, $  \tilde{c}^{(4)}_{e}\equiv \tilde{c}_4$, $ \tilde{c}^{(4)}_{(12)}\equiv\tilde{c}_3 $. From now on, we will omit the order of permutations $T_{\pi}$. The operators $ \hat{\mathcal{R}}^{(2k)}(U)$ for $k=1,2$ are evaluated explicitly in\cite{Oliviero2020random}. 

In the following, we consider scalar functions $\mathcal{P}$ that depend on $U_G\equiv e^{-i G^\dagger HG t}$. The isospectral twirling of $\mathcal{P}$ is given by
$\aver{\mathcal{P}(t)}_G=\int dG\, \mathcal{P}(G^{\dag}UG)$. 
As we shall see, if $\mathcal{P}_{\mathcal{O}}$ is characterized by a bounded operator $\mathcal{O}\in\mathcal{B}(\mathcal{H}^{\otimes 2k})$ , we obtain expressions of the form $\aver{\mathcal{P}_{\mathcal{O}}(t)}_G=\tr[\tilde{T}_\sigma \mathcal{O} \hat{\mathcal{R}}^{(2k)}(t)]$, where $\tilde{T}_\sigma$ is a normalized permutation operator, $\sigma\in S_{2k}$.

\begin{figure}[h!]
    \centering
    \includegraphics[scale=0.28]{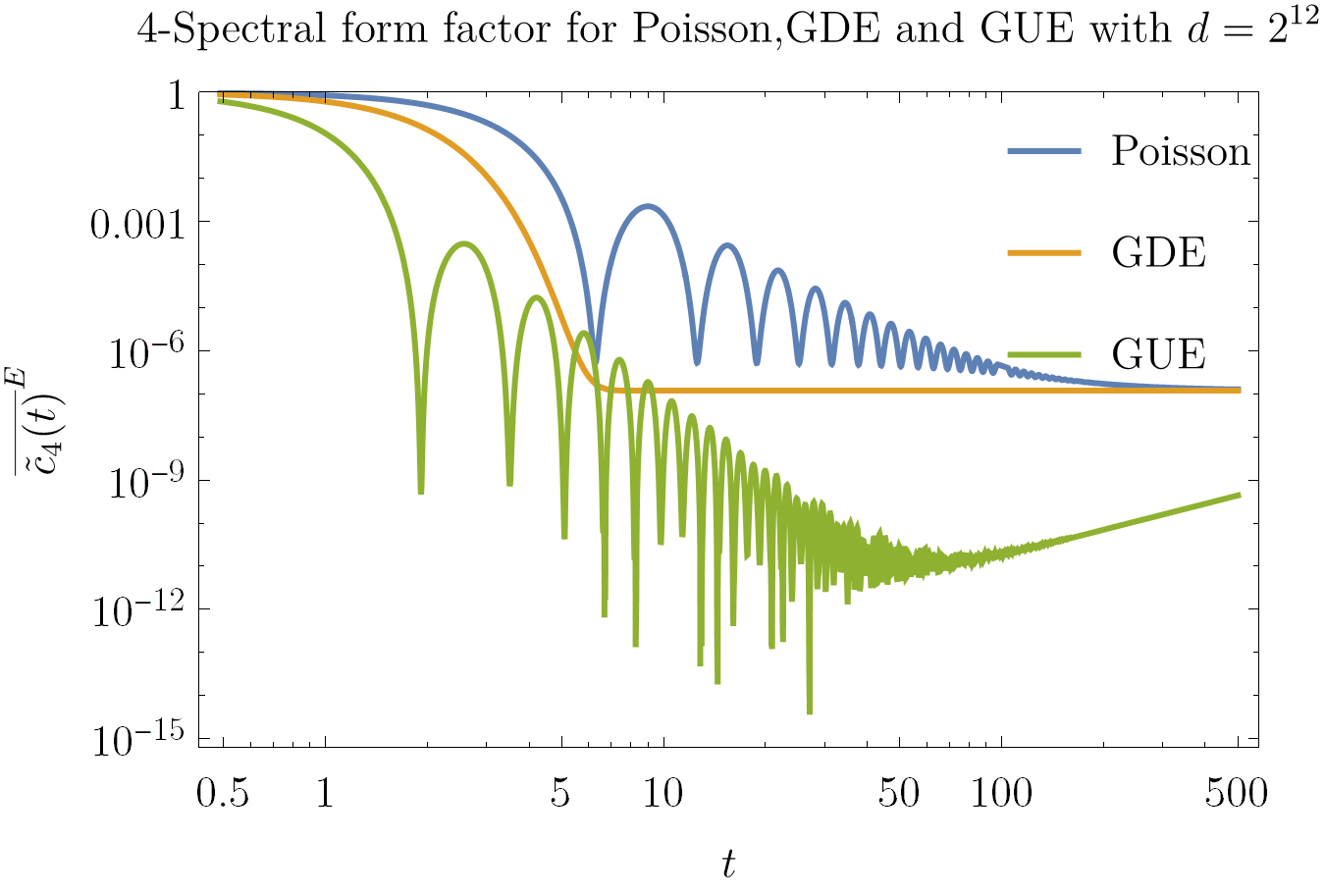}
    \caption{Log-Log plot of the spectral form factor $\overline{\tilde{c}_4(t)}^{E}$ for different ensemble E$\equiv$\text{P},  E$\equiv\text{GDE}$ and E$\equiv\text{GUE}$ for $d=2^{12}$. The starting value is $1$, while the asymptotic value is $(2d-1)d^{-3}$. }%
    \label{GUE_vs_Poisson2}%
\end{figure}

The average $\aver{\mathcal{P}_{\mathcal{O}}(t)}_G$ only depends on the spectrum of the generating Hamiltonian $H$. One can then average the value of $\aver{\mathcal{P}_{\mathcal{O}}(t)}_G$ over the spectra of an ensemble of Hamiltonians $E$. We denote such average as $\overline{\aver{\mathcal{P}_{\mathcal{O}}(t)}_{G}}^{E}$. Relevant ensembles are $E\equiv \text{GUE}$, $E\equiv \text{GDE}$ or $E\equiv \text{P}$. Since the information about the spectrum of $H$ is contained in  the $\tilde{c}_{\pi}^{(2k)}(U)$, computing $\overline{\aver{\mathcal{P}_{\mathcal{O}}(t)}_{G}}^{E}$ requires the knowledge of $\overline{\tilde{c}_{\pi}^{(2k)}(U)}^E$. The details of the random matrix calculations necessary to compute these quantities can be found in\cite{Oliviero2020random}. We present here in Fig.\ref{GUE_vs_Poisson2} the temporal evolution of  $\overline{\tilde{c}_4(t)}^{E}$, that is the most important factor for our goals. The $4-$point spectral form factor $\tilde{c}_{4}$ is able to distinguish the chaotic (GUE) and the integrable (GDE, Poisson) regime via the system-size scaling $d$. Both GUE and Poisson reach the first minimum $\overline{\tilde{c}_4(t)}^{E}=O(d^{-2})$ in a time $t=O(1)$, while GDE reach the asymptotic value $\lim_{t\rightarrow\infty}\overline{\tilde{c}_4(t)}^{\text{GDE}}=d^{-3}(2d-1)$ in a time $t=O(\sqrt{\log d})$. We observe that GUE and Poisson present a quite different temporal profile: dropping below the asymptotic value, GUE reaches the dip $\overline{\tilde{c}_4(t)}^{\text{GUE}}=O(d^{-3})$ in a time $t=O(d^{1/2})$ and then it rises to the asymptotic value $\lim_{t\rightarrow\infty}\overline{\tilde{c}_4(t)}^{\text{GUE}}=d^{-3}(2d-1)$ in a time $O(d^{-1})$; on the other hand Poisson never goes below $\lim_{t\rightarrow\infty}\overline{\tilde{c}_4(t)}^{\text{P}}=d^{-3}(2d-1)$ reaching it in a time $O(d^{1/2})$.

 In \cite{cotler2017chaos} the authors defined the twirling of the operator  $U^{\otimes k,k}$, where $U\in \mathcal{E}_{t}^{\text{GUE}}:=\{e^{-iHt}\,|\, H\in \text{\text{GUE}}\}$, i.e.
$\Phi_{\mathcal{E}_{t}^{\text{GUE}}}(U^{\otimes k,k})=\int dH\, U^{\otimes k,k} $ 
with $dH$ the unitarily invariant measure over the GUE ensemble of Hamiltonians. From $dH=d(W^{\dag}HW)$, with $W\in\mathcal{U}(\mathcal{H})$, taking the Haar average over $W$ one easily obtains:
$\Phi_{\mathcal{E}_{t}^{\text{GUE}}}(U^{\otimes t,t})=\int dH\,\hat{\mathcal{R}}^{(2k)}(U)$, i.e. the ensemble average of the isospectral twirling Eq. \eqref{isospectraltwirling} over the GUE ensemble. This approach presents some limits of applicability: unlike the isospectral twirling, it works only for a unitarily invariant distribution of Hamiltonians. In particular, it would not allow us to distinguish GUE from the integrable distributions.

%%%%%%%%%%%%%%%%%%%%%%%%%%%%%%%%%%%%%%%%%%%%%%%%
%FIGURE 1
%%%%%%%%%%%%%%%%%%%%%%%%%%%%%%%%%%%%%%%%%%%%%%%%

We now apply the isospectral twirling to probe quantum chaos. In particular, the next section is devoted to the study of the OTOCs.
\section{ OTOCs} 
Scrambling of information can be measured by two quantities, the OTOCs\cite{maldacena2016bound} and the tripartite mutual information (TMI)\cite{ding2016conditional}, namely the decay of the OTOC implies the decay of the TMI\cite{hosur2016chaos,ding2016conditional}. In this section, we show how the OTOCs are described by the isospectral twirling.
Consider $2k$ local, non-overlapping operators $A_l$, $B_l$, $l \in [1,k]$. The infinite temperature $4k-$point OTOC is defined as
\ba
\text{OTOC}_{4k}(t) &=& d^{-1}\tr(A_{1}^{\dag}(t)B_{1}^{\dag}\cdots A_{k}^{\dag}(t)B_{k}^{\dag}\nonumber\\
&\times& A_{1}(t)B_1\cdots A_{k}(t)B_{k})
\label{4kpointotoc}
\ea
where $A_l(t)=e^{iHt}A_le^{-iHt}$. Define $\mathcal{A}_{l}:=A_{l}^{\dag}\otimes A_{l}$ and similarly for $\mathcal B$. 

\begin{prop}\label{prop5}The isospectral twirling of the $4k-$point OTOC is given by
\be
\aver{\text{OTOC}_{4k}(t)}_{G}=\tr(\tilde{T}^{(4k)}_{\pi}(\otimes_{l=1}^{k}\mathcal{A}_{l}\otimes_{l=1}^{k}\mathcal{B}_{l})\hat{\mathcal{R}}^{(4k)}(U) )
\label{final4kpointotoc}
\ee
\end{prop}
See App. \ref{4kotoc} for the proof. For $k=1$ we obtain the 4-point OTOC, see App. \ref{sub4otoc}:
\be
\aver{\text{OTOC}_{4}(t)}_{G}=\tr(\tilde{T}_{(1423)}(\mathcal{A}\otimes\mathcal{B})\hat{\mathcal{R}}^{(4)}(U))
\label{final4pointotoc}
\ee
If one sets $\mathcal{A}$ and $\mathcal{B}$ to be non-overlapping Pauli operators on qubits one finds\cite{Oliviero2020random}:
\be
\aver{\text{OTOC}_{4}(t)}_{G}= \tilde{c}_4(t) -d^{-2}+O(d^{-4})
\label{4otocavergedmain}
\ee
As this result shows, the $4-$point OTOCs distinguish chaotic from integrable behavior through the timescales dictated by $\tilde{c}_4$, see Fig.\ref{GUE_vs_Poisson2}.
In a previous work\cite{cotler2017chaos} it was found $\aver{\text{OTOC}_{4}(t)}_{\text{GUE}}\simeq \tilde{c}_4(t)$; we instead remark the importance of the offset $d^{-2}$. Indeed, in \cite{Oliviero2020random} it is shown that after a time $O(d^{1/3})$ the $4-$point spectral form factor: $\overline{\tilde{c}_4(t)}^{\text{GUE}}=O(d^{-2})$; that makes the two terms in Eq. \eqref{4otocavergedmain} comparable.

\section{The role of the eigenvectors}
The Hamiltonian generating the unitary temporal evolution in a closed quantum system can be written in its spectral resolution $H=\sum_i E_i\Pi_i$, showing explicitly that the dynamics is contained both in the eigenvalues {\em and} the eigenvectors of $H$.
We have seen that, insofar only the properties of the spectrum of the Hamiltonian $H$ are concerned, different ensembles of spectra associated to different RMT distinguish the temporal profile of the chaos probes in the transient before the onset of the asymptotic behavior, which is the same for all the ensembles of spectra with a Schwartzian probability distribution\cite{Oliviero2020random}. By averaging over the unitary group in Eq.(\ref{isospectraltwirling}), we have effectively on the one hand erased any information coming from the eigenstates of the Hamiltonian, and, on the other hand, already introduced some of the properties of chaotic or ergodic Hamiltonians. For instance, these eigenvectors typically obey the eigenstate thermalization hypothesis\cite{Gemmer2010quantum,popescu2006entanglement,lloyd1988black,rigol2008thermalization}. In fact, it is striking that the spectra should have any effect at all once we use random eigenvectors. We now show that information about the asymptotic temporal behavior is contained in the way we pick the eigenvectors of $H$. To this end, 
consider a system of $N$ qubits, $\mathcal{H}=\mathbb{C}^{2^N}$, and an Hamiltonian diagonal in the computational basis $\{\ket{i}\}_{i=1}^{2^N}$, namely $H=\sum_{i}E_i\Pi_i$ with $\Pi_{i}=\ket{i}\bra{i}$ orthogonal projectors. Define then the average asymptotic unitary
\be
U^{\otimes 2,2}_{\infty}:= \lim_{t\rightarrow\infty}\overline{U^{\otimes 2,2}}^{P(E_i)}
\ee
where, as before, the average is taken over a Schwartzian probability distribution of spectra. The isospectral twirling of $U^{\otimes 2,2}_{\infty}$ will not depend on the distribution $P(E_i)$.

We now map these projectors by $\Pi_i\mapsto C\Pi_i C^\dag$ with  $C\in\mathcal C(d)$ the Clifford group. These projectors are not typical in the Hilbert space and they cannot be clearly associated to chaotic behavior, for instance it is not clear whether they feature ETH. They would possess typical entanglement but its fluctuations are not the same obtained by the Haar measure on the unitary group. Define the $Cl-$Isospectral twirling for $U^{\otimes 2,2}_{\infty}$ as
\be
\mathcal{R}_{Cl}^{(4)}(U_{\infty}):=\int_{\mathcal{C}(2^N)}\de C C^{\dag\otimes 4}U^{\otimes 2,2}_{\infty}C^{\otimes 4}
\ee
A general way to compute the Clifford average of order $4$ is to use the generalized Weingarten functions formula, which is a rearrangement of the formula showed in\cite{roth2018recovering}:
\ba
\hspace{-0.5cm}\mathcal{R}_{Cl}^{(4)}(U_{\infty})\hspace{-0.3cm}&=&\hspace{-0.3cm}\sum_{\pi\sigma}W^{+}_{g}(\pi\sigma)\tr(U^{\otimes 2,2}_{\infty}QT_{\sigma})QT_{\pi}\nonumber\hspace{0.8cm}\\\hspace{-0.3cm}&+&\hspace{-0.3cm}W^{-}_{g}(\pi\sigma)\tr(U^{\otimes 2,2}_{\infty}Q^{\perp}T_{\sigma})Q^{\perp}T_{\pi}
\ea
where $Q=\frac{1}{d^2}\sum_{P\in\mathcal{P}(2^N)}P^{\otimes 4}$, $Q^{\perp}=\bbbone^{\otimes 4}-Q$ and $P\in\mathcal{P}(2^N)$ elements of the Pauli group on $N$-qubits, while
\be
W^{\pm}_{g}(\pi\sigma):=\sum_{\lambda\,|\,D^{\pm}_{\lambda}\neq 0}\frac{d_{\lambda}^2}{(4!)^2}\frac{\chi^{\lambda}(\pi\sigma)}{D^{\pm}_\lambda}
\ee
here $\lambda$ labels the irreducible representations of the symmetric group $S_4$, $\chi^{\lambda}(\pi\sigma)$ are the character of $S_4$ depending on the irreducible representations $\lambda$, $d_\lambda$ is the dimension of the irreducible representations $\lambda$, $D_{\lambda}^{+}=\tr(QP_{\lambda})$ and $D_{\lambda}^{-}=\tr(Q^{\perp}P_{\lambda})$ where $P_\lambda$ are the projectors onto the irreducible representations of $S_{4}$, finally $T_{\sigma}$ are permutation operators corresponding to the permutation $\sigma\in S_4$. With the above formula, we compute the asymptotic value of the $4-$point OTOC:
\ba
\hspace{-0.8cm}\aver{\text{OTOC}_4}_{Cl}\!(\infty)\hspace{-0.3cm}&=&\hspace{-0.3cm}\!\sum_{\pi\sigma}W^{+}_{g}(\pi\sigma)\tr(U^{\otimes 2,2}_{\infty}QT_{\sigma})\nonumber\\\hspace{-0.3cm}&\times&\hspace{-0.3cm}\tr(\tilde{T}_{(1423)}(\mathcal{A}\otimes\mathcal{B})QT_{\pi})\nonumber\\\hspace{-0.3cm}&+&\hspace{-0.3cm}W^{-}_{g}(\pi\sigma)\tr(U^{\otimes 2,2}_{\infty}Q^{\perp}T_{\sigma})\nonumber\\\hspace{-0.3cm}&\times&\hspace{-0.3cm}\tr(\tilde{T}_{(1423)}(\mathcal{A}\otimes\mathcal{B})Q^{\perp}T_{\pi})
\ea
\begin{prop}
The asymptotic value of the $Cl-$Isospectral twirling of the $4-$point OTOC reads 
\be\label{cla}
\aver{\text{OTOC}_4}_{Cl}(\infty)=\frac{2}{(d+2)}
\ee
\end{prop}
see App. \ref{Clifford Average} for the proof. This value has to be compared with the asymptotic value for the  isospectral twirling obtained by averaging on the full unitary group:
\be\label{ua}
\aver{\text{OTOC}_4}_{G}(\infty)=\frac{1}{(d+1)(d+3)}
\ee
showing a clear separation in the asymptotic decay of the OTOCs between the full Unitary and Clifford case. For example, this shows that one cannot obtain the same asymptotic behavior by using only Clifford resources in a random quantum circuit. 

Let now the unitary evolution be generated by a $k-$doped Hamiltonian $H_k= C^{(k)\dag} H_0  C^{(k)}$ where
\ba\label{kC}
C^{(k)}= \prod_r C^\dag_r K_r
\ea
In the equation above, every $C_r\in\mathcal C(d)$ is an element of the Clifford group, while $K_r$ is a single qubit gate not belonging to the Clifford group. In this way, we have doped the Clifford Hamiltonian by non Clifford resources. Notice that, for $k=0$, the Hamiltonian is the sum of commuting Pauli strings and it is therefore integrable. If by inserting the gates $K_r$ we obtain the transition to quantum chaos, this result would also show that integrability can be deformed in a ``smooth'' way, and attain a crossover to quantum chaos. By the same technique\cite{leone2021quantum}, a lengthy but straightforward calculation gives the

\begin{theorem}
The asymptotic value of the averaged $4-$point OTOC for a $k-$diagonalizable Hamiltonian reads:
\be
\lim_{t\rightarrow\infty}\overline{\aver{\otoc_{4}(t)}_{\mathcal{C}_k}}^{P(E_i)}=\left(\frac{3}{4}\right)^{k}\frac{2}{d}+\frac{1}{d^2}+\Omega(d^{-3})
\ee
As we can see, this results interpolates between the Clifford and Haar cases of Eqs.(\ref{cla}, \ref{ua}).
\end{theorem}
As a corollary, Iff $k=\Omega(n)$, $k-$doped stabilizer Hamiltonians attain  the same scaling of \textit{Haar}  Hamiltonians for the infinite time $4-$OTOCs.

\section{Randomness of the ensemble $\mathcal{E}_H$} A chaotic Hamiltonian should generate a random unitary according to the Haar measure. To this end, we ask how random is the ensemble $\mathcal{E}_H$ 
 generated by $H$, i.e. how much the unitaries $G^{\dag}UG$ replicate the Haar distribution. We quantify randomness by computing the $k$-th frame potential of the ensemble $\mathcal{E}_H$\cite{roberts2017chaos, scott2008optimizing}, defined as
\be
\mathcal{F}_{\mathcal{E}_H}^{(k)}=\int dG_1dG_2 \left|\tr\left(G_{1}^{\dag}U^{\dag}G_{1}G_{2}^{\dag}UG_{2}\right)\right|^{2k}.
\ee
We have the following proposition:
\begin{prop}\label{prop1}The frame potential of $\mathcal{E}_H$ is the square Schatten $2-$norm of the isospectral twirling Eq. \eqref{isospectraltwirling}:
\be
\mathcal{F}_{\mathcal{E}_H}^{(k)}=\norm{\hat{\mathcal{R}}^{(2k)}(U)}_{2}^{2}=\tr\left( \tilde{T}_{1\leftrightarrow2}\left(\hat{\mathcal{R}}^{\dag(2k)}\otimes  \hat{\mathcal{R}}^{(2k)}\right)\right)
\label{framepoth}
\ee
where $\tilde{T}_{1\leftrightarrow2}$ is the swap operator between the first $2k$ copies of $\mathcal H$ and the second $2k$ copies.
\end{prop}
See App. \ref{framepot1} for the proof. The Haar value $\mathcal{F}_{\text{Haar}}^{(k)}=k!$ is a lower bound to this quantity\cite{roberts2017chaos}, that is,
$\mathcal{F}_{\text{Haar}}^{(k)}\le \mathcal{F}_{\mathcal{E}_H}^{(k)}$ 
so a larger value of the frame potential means less  randomness. 

\begin{prop}\label{prop2} The frame potential of $\mathcal{E}_H$ obeys the following lower bound:
\be
\mathcal{F}_{\mathcal{E}_H}^{(k)} \ge d^{-2k}\left|\tr(U)\right|^{4k}
\label{bound1}
\ee
\end{prop}
The above result is useful to see if an ensemble deviates from the Haar distribution, see App. \ref{framepot2} for the proof. Taking the infinite time average $\mathbb{E}_T (\cdot)=\lim_{T\to\infty}T^{-1}\int^T_0(\cdot)dt$ of the r.h.s, we can calculate a lower bound for the asymptotic value of the frame potential.

\begin{figure}
    \centering
   	\includegraphics[scale=0.28]{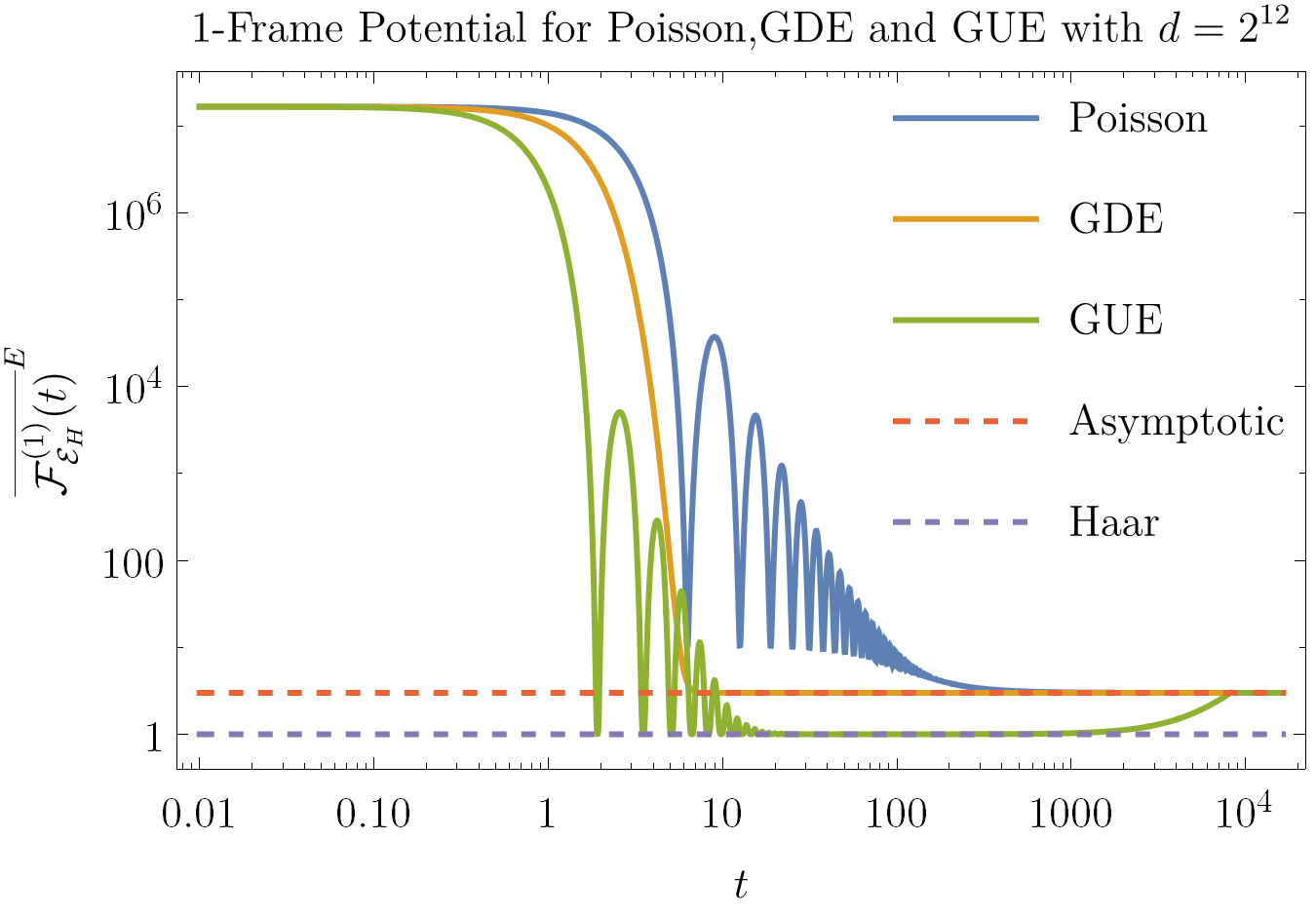} 
   \caption{Log-Log plot of the ensemble average of $\overline{\mathcal{F}^{(1)}_{\mathcal{E}_H}(t)}^{E}$ for $E\equiv\text{GUE}$, $E\equiv\text{GDE}$ and $E\equiv\text{P}$ for $d=2^{12}$. The dashed lines represent the Haar value  $\mathcal{F}_{\text{Haar}}^{(1)}=1$ and the asymptotic value of $\lim_{t\rightarrow\infty}\overline{\mathcal{F}^{(1)}_{\mathcal{E}_H}(t)}^{E}=3+O(d^{-1})$. Note that at late times $t=O(d)$,  $\overline{\mathcal{F}_{\mathcal{E}_{H}}^{(1)}}^{\text{GUE}}$  distances from the Haar value\cite{cotler2017chaos} and reaches the asymptotic value.
   }
   \label{fig:Framepotential}
\end{figure}

\begin{prop}\label{prop3}If the spectrum of $H$ is {\em generic}:
\be
\mathbb{E}_T\left[d^{-2k}\left|\tr(U)\right|^{4k}\right]=(2k)!+O(d^{-1})
\label{timeaveframe}
\ee
\end{prop}
As we can see, it is far from the Haar value $k!$. The request for the spectrum being generic is a stronger form of non resonance, see App. \ref{subsecgeneric} for the definition of generic spectrum and App. \ref{framepot3}  for the proof. The infinite time average shows that the asymptotic value is the same for GUE and GDE. On the other hand, the frame potential $\mathcal{F}_{\mathcal{E}_H}^{(k)}$ is non trivial in its time evolution.  For $k=1$, we have\cite{Oliviero2020random}:
\be
\mathcal{F}_{\mathcal{E}_H}^{(1)}=\frac{d^2}{(d^2-1)}(d^2\tilde{c}_4(t)-2\tilde{c}_2(t)+1)
\ee
where of course the coefficients $\tilde{c}_k(t)$ do depend on the spectrum of $H$. We 
can now take the ensemble average $\overline{\mathcal{F}_{\mathcal{E}_H}^{(1)}}^E$ of this quantity. The results are plotted in Fig.\ref{fig:Framepotential}. We can see that the behavior of the Poisson and GDE spectra is quite distinct from that of the GUE. Indeed, for the first two ensembles, the frame potential
never goes below the asymptotic value $3+O(d^{-1})$, so it always stays away from the Haar value $1$. On the other hand, the frame potential, corresponding to GUE, equals the Haar value $1$ in the whole temporal interval $t\in [O(d^{1/3}),O(d)]$.

%%%%%%%%%%%%%%%%%%%%%%%%%%%%%%%%%%%%%%%%%%%%%%%%
%FIGURE 2
%%%%%%%%%%%%%%%%%%%%%%%%%%%%%%%%%%%%%%%%%%%%%%%%
%%%%%%%%%%%%%%%%%%%%%%%%%%%%%%%%%%%%%%%%%%%%%%%%
%FIGURE 2
%%%%%%%%%%%%%%%%%%%%%%%%%%%%%%%%%%%%%%%%%%%%%%%%

\begin{prop}\label{prop4}The ensemble average of the frame potential for $E\equiv \text{GDE}$ satisfies: 
\be
\overline{\mathcal{F}_{\mathcal{E}_H}^{(k)}}^{\text{GDE}}\ge (2k)!+O(d^{-1})
\label{GDEbehavior}
\ee
showing that the GDE ensemble is always different from the Haar value.
\end{prop}
See App. \ref{framepot4} for the details of the proof.

\section{ Loschmidt Echo and OTOC}
The Loschmidt Echo (LE) is a quantity that captures the sensitivity of the dynamics to small perturbations. 
In\cite{yan2020information,bhattacharyya2019web}, it was found that, under suitable conditions, the OTOC and LE are quantitatively equivalent. Our aim in this section is to give another insight in that direction, showing that, using the isospectral twirling, the LE assumes the form of an OTOC-like quantity. The LE is defined as $\mathcal L(t)=d^{-2}|\tr(e^{iHt}e^{-i(H+\delta H)t})|^2$.

\begin{prop}\label{prop6} Let $A\in\mathcal{U}(\mathcal{H})$ be a unitary operator, provided that $\text{Sp}(H)=\text{Sp}(H+\delta H)$ the isospectral twirling of the LE is given by: 
\ba
\aver{\mathcal{L}(t)}_{G}=\tr(\tilde{T}_{(14)(23)}\mathcal{A}^{\otimes 2}\hat{\mathcal{R}}^{(4)}(U))
\label{loschmfinal}
\ea
\end{prop}
where $\mathcal{A}:=A^\dag\otimes A$. See App. \ref{le10} for the proof. If one sets $\mathcal{A}$ to be a Pauli operator on qubits one gets\cite{Oliviero2020random}:
\be
\aver{\mathcal{L}(t)}_{G}= \tilde{c}_4(t) +d^{-2}+O(d^{-4})
\ee
In conclusion, we can say that both LE and OTOC are proportional to the $4-$point spectral form factor in this setting. We can conclude that also the LE is a probe of scrambling; we thus find an agreement with the statement of\cite{cotler2017chaos}. Indeed, in proving Eq. \eqref{loschmfinal}, we give an expression of the LE in terms of the $2-$point auto-correlation function $|\tr(A^{\dag}(t)A)|^{2}$; in \cite{cotler2017chaos} it was proved that the decay of the averaged $2-$point autocorrelation function implies the decay of the TMI, i.e. implies scrambling \cite{hosur2016chaos,ding2016conditional}.

\section{Entanglement}
We now move onto showing how the isospectral twirling also describes the evolution of entanglement under a random Hamiltonian with a given spectrum. 
Consider the unitary time evolution of a state $\psi \in \mathcal B(\mathcal H_A \otimes\mathcal H_B)$ by $\psi\mapsto \psi_t\equiv U\psi U^\dagger$. The entanglement of $\psi_t$ in the given bipartition is computed by the $2-$R\'enyi entropy $S_2 =-\log \tr (\psi_A(t)^2) $, where $\psi_A(t):=\tr_B\psi_t$.

\begin{prop}\label{prop7}
The isospectral twirling of the $2-$Renyi entropy is lower bounded by:
\be
\hspace{-0.4cm}\aver{S_{2}}_{G} \ge -\log\tr\left(T_{(13)(24)}\hat{\mathcal R}^{(4)}(U)\psi^{\otimes 2} \otimes T_{(A)}\right)
\label{finalentanglement}
\ee
where $T_{(A)}\equiv T_{A}\otimes \bbbone_{B}^{\otimes 2}$ and $T_A$ is the swap operator on $\mathcal{H}_{A}$.
\end{prop}
See App. \ref{subenta} for the proof. If one sets $d_{A}=d_{B}=\sqrt{d}$, one gets\cite{Oliviero2020random}:
\ba
\hspace{-0.4cm}\aver{S_{2}}_{ G}\hspace{-0.3cm}&\ge&\hspace{-0.3cm}-\log\left[2d^{-1/2}+\tilde{c_{4}}(t)\left(\tr(\psi_{A}^2)-2d^{-1/2}\right)\right]\nonumber\hspace{-1cm}\\&+&\hspace{-0.3cm}O(1/d)
\ea
As the temporal behavior of $\aver{S_{2}}_{ G}$ is dictated by $\tilde{c_{4}}(t)$, one expects that also entanglement  dynamics can distinguish between chaotic and non-chaotic behavior. The complete analysis of this dynamics is to be found in\cite{Oliviero2020random}.

\section{Tripartite mutual information} The TMI is defined as \cite{ding2016conditional,hosur2016chaos} 
$
I_{3}(A:C:D):=I(A:C)+I(A:D)-I(A:CD)
$ 
where $A,B$ and $C,D$ are fixed bipartitions of past and future time slices of the quantum system after a unitary evolution $U$; $I(A:C)$ is the mutual information defined through the Von Neumann entropy. Here we work with the TMI using the 2-R\'enyi entropy as measure of entropy, and denote it by  $I_{3_{(2)}}(U)=\log d +\log\tr\rho_{AC}^{2}+\log\tr\rho_{AD}^{2}$ see App. \ref{TMI}. Here $\rho_{AC(AD)}=\tr_{BD(BC)}(\rho_U)$, where $\rho_U$ is the Choi state\cite{choi1975completely} of the unitary evolution $U\equiv\exp\{-iHt\}$ and $H$ a random Hamiltonian with a given spectrum.  Set $A=C$ and $B=D$, then, by defining $T_{(C)}^{U}:=U^{\otimes 2}T_{(C)}U^{\dag\otimes 2}$,  $I_{3_{(2)}}$ can be written as (see App. \ref{prooftracedTMI}):
\be
I_{3_{(2)}}\!\!=\!\!-3\log d+\log\tr (T_{(C)}^{U}T_{(C)}) +  \log\tr (T_{(C)}^{U}T_{(D)})
\label{tracedTMI}
\ee
The second term of Eq. \eqref{tracedTMI} reminds of the entanglement of quantum evolutions defined in\cite{zanardi2001entanglement}.
 
\begin{figure}
    \centering
    \includegraphics[scale=0.28]{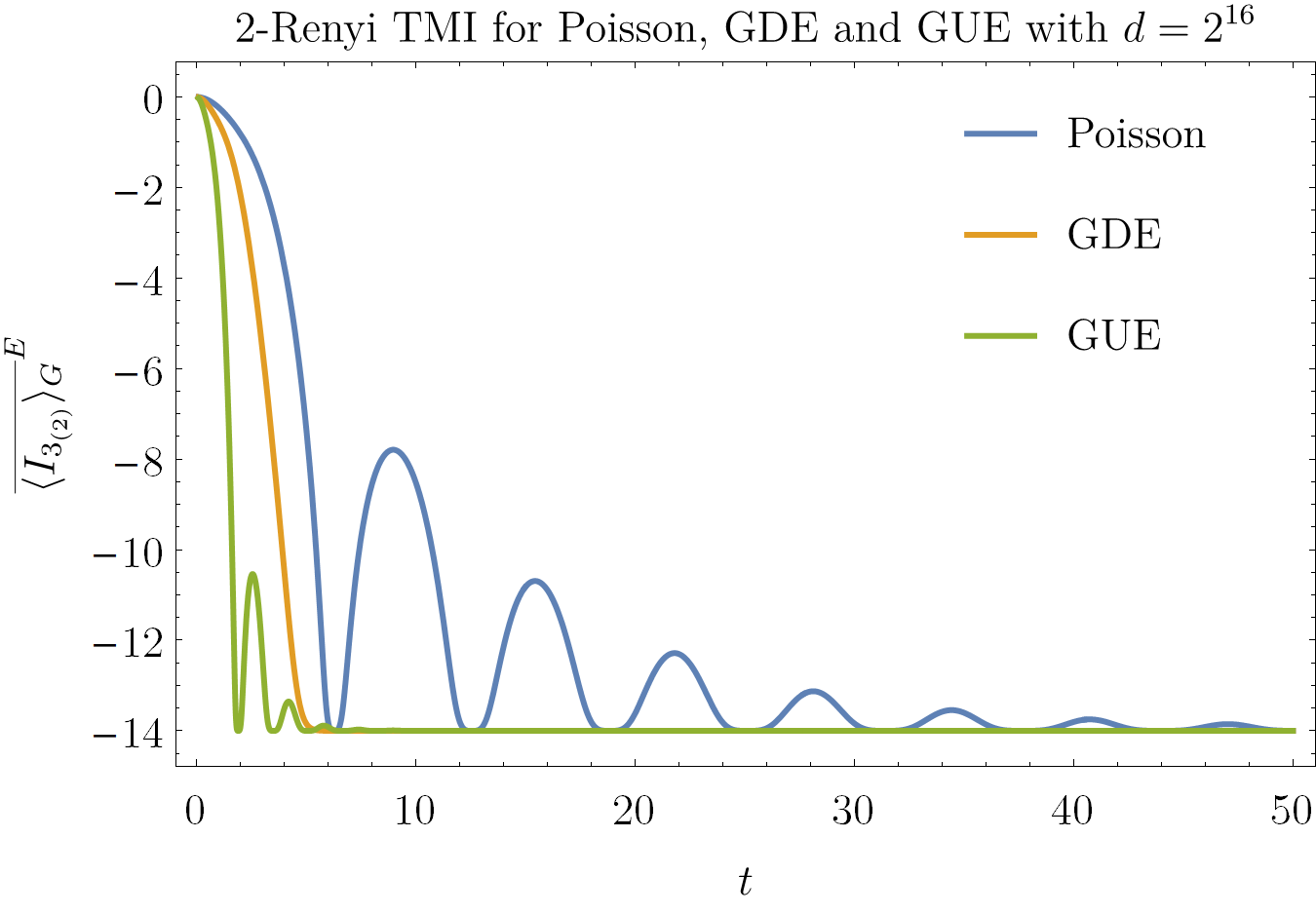}
    \caption{Plot of the upper bound for $\overline{\aver{I_{3_{(2)}}(t)}_{G}}^{E}$, see r.h.s of Eq. \eqref{I3r4}, for E$\equiv$P, E$\equiv$GUE and E$\equiv$GDE  with $d_C=d_D=d^{1/2}$ and $d=2^{16}$. GUE and Poisson reveal oscillations before the plateau whose amplitude and damp time increases with the system size $d$, see Eq. \eqref{plateaudcsqrtd} and Eq. \eqref{behavior} respectively.  For GDE there are no oscillations: the plateau is reached in $O(\sqrt{\log d})$.}%
    \label{GUE_vs_Poisson4}%
\end{figure}

\begin{prop}\label{prop8} The isospectral twirling of $I_{3_{(2)}}$ is upper bounded by:
\ba
\hspace{-0.3cm}\aver{I_{3_{(2)}}}_{G}\hspace{-0.3cm}&\le&\hspace{-0.3cm} \log d +\log\tr (\tilde{T}_{(13)(24)}\hat{\mathcal{R}}^{(4)}(U) T_{(C)}^{\otimes 2})
\nonumber\hspace{0.8 cm}\\\hspace{-0.3cm}&+&\hspace{-0.3cm}\log\tr (\tilde{T}_{(13)(24)}\hat{\mathcal{R}}^{(4)}(U)T_{(C)}\otimes T_{(D)})\label{I3r4}
\ea
\end{prop}
Since the TMI is a negative-definite quantity, the decay of the r.h.s. of Eq. \eqref{I3r4} implies scrambling, i.e. the l.h.s. drop closer to its minimum value. The tightness of this bound deserves further investigations.
By computing explicitly Eq.\ref{I3r4} one has\cite{Oliviero2020random}:
\ba
\hspace{-0.3cm}\aver{ I_{3_{(2)}}(t)}_{ G}\hspace{-0.3cm}&\le&\hspace{-0.3cm}\log_2 (2-3\tilde{c}_4(t) + 2 \text{Re}\tilde{c}_3(t))\label{2rtmisqrtd}\\\hspace{-0.3cm}&+&\hspace{-0.3cm}\log_2(\tilde{c}_4(t)+(2-\tilde{c}_4(t))d^{-1})+O(d^{-2})\nonumber
\ea
We can now compute $\overline{\aver{I_{3_{(2)}}}_{G}}^E$ over the spectra GUE, GDE and Poisson. We set $d_C=\dim(\mathcal{H}_C)$ and $d_{D}=\dim\mathcal{H}_D$ and $d_{C}=d_{D}=\sqrt{d}$. The time evolution of $\overline{\aver{I_{3_{(2)}}}_{G}}^E$ depends on the spectral form factors. We can see in Fig.\ref{GUE_vs_Poisson4} how the chaotic and integrable behaviors are clearly different. 
The salient timescales of $I_{3_{(2)}}(t)$ depend on the timescales of $\tilde{c}_4(t)$\cite{Oliviero2020random}. 
The plateau value of Eq. \eqref{2rtmisqrtd} are, for large $d$:
\be
\lim_{t\rightarrow \infty}\overline{\aver{I_{3_{(2)}}(t)}_{G}}^{E}=2-\log _{2} d+O(d^{-1})
\label{plateaudcsqrtd} 
\ee
%%%%%%%%%%%%%%%%%%%%%%%%%%%%%%%%%%%%%%%%%%%%%%%%
%FIGURE 3
%%%%%%%%%%%%%%%%%%%%%%%%%%%%%%%%%%%%%%%%%%%%%%%%
%%%%%%%%%%%%%%%%%%%%%%%%%%%%%%%%%%%%%%%%%%%%%%%%
%FIGURE 3
%%%%%%%%%%%%%%%%%%%%%%%%%%%%%%%%%%%%%%%%%%%%%%%%
One thing to note  is that the fluctuations of GUE and Poisson decay in a time
\be
t_{\text{fluct}}=\alpha+ \beta \log d
\label{behavior}
\ee
where the parameters $\alpha, \beta$ for the different ensembles  are GUE: $\alpha=-3.9,\beta=0.8$ Poisson  $\alpha=-16.3,\beta=3.2$.

\section{ Conclusions and Outlook} Chaos is an important subject in quantum many-body physics and the understanding of its appearance is of fundamental importance for a number of situations ranging from quantum information algorithms to black hole physics. In this paper, we unified the plethora of probes to quantum chaos in the notion of isospectral twirling. Since this quantity depends explicitly on the spectrum of the Hamiltonian, one can compare its behavior for different spectra characterizing chaotic and non-chaotic behavior, which we did by using random matrix theory. We demonstrate how different temporal features depend on the interplay between spectrum and eigenvectors of the Hamiltonian. Random eigenvectors obtained with Clifford resources result in markedly different asymptotic values of the OTOCs. Moreover, a doping of Clifford circuits by non-Clifford resources interpolates the long time scaling of the OTOCs between a class of integrable models and quantum chaos.  

In perspective, there are several open questions. First and foremost, we want to extend the results of the crossover to more structural aspects of the dynamics with the goal of obtaining a quantum KAM theorem. Second, one could systematically study how different spectra behave together with different ensembles of eigenvectors, for instance interpolating between Clifford and universal resources in a random quantum circuit\cite{zhou2020single}, by doping a stabilizer Hamiltonian with non Clifford resources like the $T-$gates\cite{leone2021quantum}.  Another important aspect is that of the locality of the interactions. In this work, we did not take into account the locality of interactions. Locality might result in even more striking differences in the onset of quantum chaotic behavior. In this paper, we have treated the spectrum and the eigenvectors of the Hamiltonian separately, showing how they both contribute to quantum chaotic features. This is possible because in the spectral resolution spectrum and eigenvectors are distinct. However, in realistic systems, we often find that {\em both spectra and eigenvectors} possess quantum chaotic features; this should depend on the fact that we deal with local Hamiltonians. 
Through the connections found with entanglement and quantum thermodynamics, one also hopes to exploit these findings to design more efficient quantum batteries. Finally,  the notion of isospectral twirling could be generalized to non unitary quantum channels and used to study chaotic behavior in open quantum systems.

\section{Acknowledgments} We acknowledge  support from  NSF award number 2014000. We thank F. Caravelli for enlightening discussions.

%%%%%%%%%%%%%%%%%%%%%%%%%%%%%%%%%%%%%%%%%%%%%%%%%%%%%
\appendix

\section{$4k-$point OTOC\label{4kotoc}}
\subsection{Proof of Proposition \ref{prop5}}
Recall the definition $(6)$ and use the cyclic property of the trace to write it in terms of $B_{l}(-t)=UB_l U^{\dag}$:
\ba
\hspace{-0.6cm}\text{OTOC}_{4k}=\frac{1}{d}\tr(&&\hspace{-0.7cm}A_{1}^{\dag}B_{1}^{\dag}(-t)\cdots A_{k}^{\dag}B_{k}^{\dag}(-t)\times\nonumber \\&&\hspace{-0.7cm}A_{1}B_1(-t)\cdots A_{k}B_{k}(-t))
\ea
to write the product of operators in terms of a tensor product, let us use the property proven in \cite{Oliviero2020random}:
\ba
\hspace{-0.6cm}\text{OTOC}_{4k}=\tr(\tilde{T}_{1\,4k\cdots 2}&&\hspace{-0.7cm}(A_{1}^{\dag}\otimes B_{1}^{\dag} \otimes \cdots\otimes A_{k}\otimes B_k )\times\nonumber\\&&\hspace{-1.4cm}(U\otimes U^{\dag})^{\otimes 2k})
\ea
where $(1\,4k\,\cdots 2)\in S_{4k}$. 
Now let us act with the adjoint action of $S:= \prod_{l=0}^{k-1}T_{(2l+2\,\,2k+2l+1)}$, $(2l+2\,\,2k+2l+1)\in S_{4k}$ on $(U\otimes U^{\dag})^{\otimes 2k}$; inserting multiple $S^{\dag}S\equiv \bbbone$ one gets:
\be
\text{OTOC}_{4k}=\tr\left(\tilde{T}^{(4k)}_{\pi}(A_{1}^{\dag}\otimes A_{1} \otimes \cdots\otimes B_{k}^{\dag}\otimes B_k )U^{\otimes k,k}\right)\label{aotoc4k}
\ee
where
$\tilde{T}^{(4k)}_{\pi}=S\tilde{T}_{1\,4k\cdots 2}S^{\dag}$; defining $\mathcal{A}_l=\otimes_{l=1}^{k}A_{l}^{\dag}\otimes A_l$, similarly for $\mathcal{B}$ and averaging over $U_{G}:=G^{\dag}UG$ one obtains the desired result $(7)$.
\subsection{Proof of Eq. $\eqref{final4pointotoc}$}\label{sub4otoc}
Setting $k=1$, we have ${T}^{(4)}_{\pi}=T_{(23)}T_{(1432)}T_{(23)}=T_{(1423)}$, therefore Eq. \eqref{aotoc4k} after the isospectral twirling reads:
\be
\text{OTOC}_{4}=\tr\left(T_{(1423)}(\mathcal{A}\otimes\mathcal{B})R^{(4)}(U)\right)
\ee
this concludes the proof.
\section{Calculations for the Clifford averages}\label{Clifford Average}
\subsection{Calculation of $\tr(U^{\otimes 2,2}_{\infty}QT_{\sigma})$}
Defining $\Pi_{ijij}\equiv\Pi_{i}\otimes \Pi_{j}\otimes \Pi_{i}\otimes \Pi_{j}$ and similarly for the others, $U^{\otimes 2,2}_{\infty}$ reads:
\be
U^{\otimes 2,2}_{\infty}=\sum_{i\neq j}(\Pi_{ijij}+\Pi_{ijji})+\sum_{i}\Pi_{iiii}
\label{Uinfty}
\ee
Then since let us split $Q$ into two parts:
\be
Q=\frac{\bbbone^{\otimes 4}}{d^2}+\sum_{P\neq \bbbone}P^{\otimes 4}\equiv \frac{\bbbone^{\otimes 4}}{d^2}+Q^{\prime}
\ee
We write:
\be
\tr(U^{\otimes 2,2}_{\infty}QT_{\sigma})=\frac{1}{d^2}\tr(U^{\otimes 2,2}T_{\sigma})+\tr(U^{\otimes 2,2}_{\infty}Q^{\prime}T_{\sigma})
\ee
and note that the first part coincides with the asymptotic values of these traces used for the computation of the usual Isospectral twirling, see \cite{Oliviero2020random}:
\be
\tr(U^{\otimes 2,2}_{\infty}QT_{\sigma})=\frac{1}{d^2}\tr(U^{\otimes 2,2}T_{\sigma})+\tr(U^{\otimes 2,2}_{\infty}Q^{\prime}T_{\sigma})
\ee
the only parts which is left out and need an evaluation is $\tr(U^{\otimes 2,2}_{\infty}Q^{\prime}T_{\sigma})$, with $Q^{\prime}=\sum_{P\neq \bbbone}P^{\otimes 4}$
Therefore plugging Eq. \eqref{Uinfty} we find:
\ba
d^2\tr(U_{\infty}^{\otimes 2,2}Q^{\prime})\hspace{-0.3cm}&=&\hspace{-0.3cm}2d(d-1)^2+d(d-1)\nonumber\\
d^2\tr(U_{\infty}^{\otimes 2,2}Q^{\prime}T_{(12)})\hspace{-0.3cm}&=&\hspace{-0.3cm}\tr(U_{\infty}^{\otimes 2,2}Q^{\prime}T_{(34)})\nonumber\\
\hspace{-0.3cm}&=&\hspace{-0.3cm}d(d-1)\\
d^2\tr(U_{\infty}^{\otimes 2,2}Q^{\prime}T_{(23)})\hspace{-0.3cm}&=&\hspace{-0.3cm}\tr(U_{\infty}^{\otimes 2,2}Q^{\prime}T_{(24)})\nonumber\\
\hspace{-0.3cm}&=&\hspace{-0.3cm}\tr(U_{\infty}^{\otimes 2,2}Q^{\prime}T_{(13)})\nonumber\\\hspace{-0.3cm}&=&\hspace{-0.3cm}\tr(U_{\infty}^{\otimes 2,2}Q^{\prime}T_{(14)})\nonumber\\\hspace{-0.3cm}&=&\hspace{-0.3cm}d(d-1)^2+d(d-1)\nonumber\\
d^2\tr(U_{\infty}^{\otimes 2,2}Q^{\prime}T_{(ijk)})\hspace{-0.3cm}&=&\hspace{-0.3cm}d(d-1)\nonumber\\
d^2\tr(U_{\infty}^{\otimes 2,2}Q^{\prime}T_{(1234)})\hspace{-0.3cm}&=&\hspace{-0.3cm}\tr(U_{\infty}^{\otimes 2,2}Q^{\prime}T_{(1432)})\nonumber\\\hspace{-0.3cm}&=&\hspace{-0.3cm}\tr(U_{\infty}^{\otimes 2,2}Q^{\prime}T_{(1243)})\nonumber\\\hspace{-0.3cm}&=&\hspace{-0.3cm}\tr(U_{\infty}^{\otimes 2,2}Q^{\prime}T_{(1342)})\nonumber\\\hspace{-0.3cm}&=&\hspace{-0.3cm}d^2(d-1)+d(d-1)\nonumber\\
d^2\tr(U_{\infty}^{\otimes 2,2}Q^{\prime}T_{(1324)})\hspace{-0.3cm}&=&\hspace{-0.3cm}\tr(U_{\infty}^{\otimes 2,2}Q^{\prime}T_{(1423)})\nonumber\\\hspace{-0.3cm}&=&\hspace{-0.3cm}d(d-1)\nonumber\\
d^2\tr(U_{\infty}^{\otimes 2,2}Q^{\prime}T_{(12)(34)})\hspace{-0.3cm}&=&\hspace{-0.3cm}2d^2(d-1)+d(d-1)\nonumber\\
d^2\tr(U_{\infty}^{\otimes 2,2}Q^{\prime}T_{(13)(24)})\hspace{-0.3cm}&=&\hspace{-0.3cm}\tr(U_{\infty}^{\otimes 2,2}Q^{\prime}T_{(14)(23)})\nonumber\\\hspace{-0.3cm}&=&\hspace{-0.3cm}d^2(d-1)+d(d-1)^2\nonumber\\\hspace{-0.3cm}&+&\hspace{-0.3cm}d(d-1)\nonumber
\ea
The above calculations are straightforward, let us just give some insights and examples. For $\Pi_{iiii}$ the result is always the same, indeed $T_{\sigma}\Pi_{iiii}=\Pi_{iiii}$ and:
\ba
d^2\sum_{i}\tr(\Pi_{iiii}Q)&=&\sum_{i,P\neq \bbbone}|\langle i|P|i\rangle|^{4}\\\hspace{-0.3cm}&=&\hspace{-0.3cm}\sum_{i}\sum_{P\ni \{Z,I\}\neq \bbbone}=d(d-1)\nonumber
\ea
Or let us calculate $\sum_{i\neq j}\tr(\Pi_{ijij}QT_{(1234)})$:
\ba
d^2\sum_{i\neq j}\tr(\Pi_{ijij}QT_{(1234)})\hspace{-0.3cm}&=&\hspace{-0.3cm}\sum_{i\neq j,P\neq \bbbone}\langle ijij|P|jiji\rangle\nonumber\hspace{0.4cm}\\\hspace{-0.3cm}&=&\hspace{-0.3cm}\sum_{i\neq j,P\neq \bbbone}|\langle i|P|j\rangle |^{4}\\\hspace{-0.3cm}&=&\hspace{-0.3cm}\sum_{i\neq j, P\neq \bbbone }=d^2(d-1)\nonumber
\ea
Indeed for any pair of $(i,j)$ with $i\neq j$ there are $d$ Pauli operators which transforms $i$ into $j$ and viceversa. While for $Q^{\perp}$ we have:
\be
\tr(U^{\otimes 2,2}_{\infty}Q^{\perp}T_{\sigma})=\frac{d^2-1}{d^2}\tr(U^{\otimes 2,2}T_{\sigma}) -\tr(U^{\otimes 2,2}_{\infty}Q^{\prime}T_{\sigma})
\ee
\subsection{Calculation of $\tr(\tilde{T}_{(1423)}\mathcal{A}\otimes\mathcal{B}QT_{\pi})$}
Let us calculate this trace for all $\pi\in S_4$:
\ba
\tr(\tilde{T}_{(1423)}\mathcal{A}\otimes\mathcal{B}Q\bbbone)\hspace{-0.3cm}&=&\hspace{-0.3cm}\tr(\tilde{T}_{(1423)}\mathcal{A}\otimes\mathcal{B}QT_{(ij)(kl)})\nonumber\\\hspace{-0.3cm}&=&\hspace{-0.3cm}d^{-1}\tr(ABAB)=1\\
\tr(\tilde{T}_{(1423)}\mathcal{A}\otimes\mathcal{B}QT_{(12)})\hspace{-0.3cm}&=&\hspace{-0.3cm}\tr(\tilde{T}_{(1423)}\mathcal{A}\otimes\mathcal{B}QT_{(34)})\nonumber\\\hspace{-0.3cm}&=&\hspace{-0.3cm}\tr(\tilde{T}_{(1423)}\mathcal{A}\otimes\mathcal{B}QT_{(ijk)})\nonumber\\\hspace{-0.3cm}&=&\hspace{-0.3cm}\tr(\tilde{T}_{(1423)}\mathcal{A}\otimes\mathcal{B}QT_{(1324)})\nonumber\\\hspace{-0.3cm}&=&\hspace{-0.3cm}\tr(\tilde{T}_{(1423)}\mathcal{A}\otimes\mathcal{B}QT_{(1423)})=0\nonumber\\
\tr(\tilde{T}_{(1423)}\mathcal{A}\otimes\mathcal{B}QT_{(13)})\hspace{-0.3cm}&=&\hspace{-0.3cm}\tr(\tilde{T}_{(1423)}\mathcal{A}\otimes\mathcal{B}QT_{(23)})\nonumber\\\hspace{-0.3cm}&=&\hspace{-0.3cm}\tr(\tilde{T}_{(1423)}\mathcal{A}\otimes\mathcal{B}QT_{(14)})\nonumber\\\hspace{-0.3cm}&=&\hspace{-0.3cm}\tr(\tilde{T}_{(1423)}\mathcal{A}\otimes\mathcal{B}QT_{(24)})\nonumber\\\hspace{-0.3cm}&=&\hspace{-0.3cm}\tr(\tilde{T}_{(1423)}\mathcal{A}\otimes\mathcal{B}QT_{(1234)})\nonumber\\\hspace{-0.3cm}&=&\hspace{-0.3cm}\tr(\tilde{T}_{(1423)}\mathcal{A}\otimes\mathcal{B}QT_{(1342)})\nonumber\\\hspace{-0.3cm}&=&\hspace{-0.3cm}\tr(\tilde{T}_{(1423)}\mathcal{A}\otimes\mathcal{B}QT_{(1243)})\nonumber\\\hspace{-0.3cm}&=&\hspace{-0.3cm}\tr(\tilde{T}_{(1423)}\mathcal{A}\otimes\mathcal{B}QT_{(1432)})\nonumber\\\hspace{-0.3cm}&=&\hspace{-0.3cm}d^{-2}\tr(ABAB)=d^{-1}\nonumber
\ea
the above calculations are straightforward, let us discuss just one example:
\ba
\tr(\tilde{T}_{(1423)}\mathcal{A}\otimes\mathcal{B}QT_{(1324)})\hspace{-0.3cm}&=&\hspace{-0.3cm}d^{-1}\tr(\mathcal{A}\otimes \mathcal{B}Q)\nonumber\\\hspace{-0.3cm}&=&\hspace{-0.3cm}d^{-3}\sum_{P}\tr(AP)^2\tr(BP)^2\nonumber\\\hspace{-0.3cm}&=&\hspace{-0.3cm}d\sum_{P}\delta_{AP}\delta_{BP}=0
\ea
since $A$ and $B$ are non-overlapping Pauli operators.
\section{Frame potential\label{framepot}}
\subsection{Proof of Proposition \ref{prop1}}\label{framepot1}
Recall the definition of the frame potential
\be
\mathcal{F}_{\mathcal{E}_H}^{(k)}=\int dG_1dG_2 \left|\tr\left(G_{1}^{\dag}U^{\dag}G_{1}G_{2}^{\dag}UG_{2}\right)\right|^{2k}
\ee
Then:
\ba
\mathcal{F}_{\mathcal{E}_H}^{(k)}&=&\int dG_1dG_2\tr\left(G_{1}^{\dag}UG_{1}G_{2}^{\dag}U^{\dag}G_{2}\right)^k\nonumber\\
&\times&\tr\left(G_{1}^{\dag}U^{\dag}G_{1}G_{2}^{\dag}UG_{2}\right)^k
\ea
Using the property of the trace for which $\tr(A)^{k}=\tr(A^{\otimes k})$ we can rewrite it as:
\be
\hspace{-0.3cm}\mathcal{F}_{\mathcal{E}_H}^{(k)}=\!\!\int dG_1dG_2\tr\left(G_{1}^{\dag\otimes 2k}U^{\dag\otimes k,k}G_{1}^{\otimes 2k}G_{2}^{\dag\otimes 2k}U^{\otimes k,k}G_{2}^{\otimes 2k}\right)
\ee
where $U^{\otimes k,k}\equiv U^{\otimes k}\otimes U^{\dag\otimes k}$. From the definition $(1)$ we have:
\be
\mathcal{F}_{\mathcal{E}_H}^{(k)}=\tr\left(\hat{\mathcal{R}}^{(2k)^\dag}(U)\hat{\mathcal{R}}^{(2k)}(U)\right)=\norm{\hat{\mathcal{R}}^{(2k)}(U)}^{2}_{2}
\ee
the result is proven. To write $\mathcal{F}_{\mathcal{E}_H}^{(k)}$ in a linear form, define ${T}_{1\leftrightarrow2}\equiv \prod_{l=1}^{2k}T_{(l\,2k+l)}$:
\be
\mathcal{F}_{\mathcal{E}_H}^{(k)}=\tr\left({T}_{1\leftrightarrow2}\left( \hat{\mathcal{R}}^{(2k)^\dag}\otimes \hat{\mathcal{R}}^{(2k)}\right)\right)
\ee
\subsection{Proof of Proposition \ref{prop2}}\label{framepot2}
In order to prove Proposition \ref{prop2} we make use the usual bound holding for the Schatten p-norms\cite{watrous2018theory}
$\norm{A}_{p}\le \text{rank}(A)^{\frac{1}{p}-\frac{1}{q}}\norm{A}_q$. Since $\text{rank}(A)=\text{rank}(A^{\dag}A)$, we have $\text{rank}(\hat{\mathcal{R}}^{(2k)}(U))=d^{2k}$ and thus:
\be
\norm{\hat{\mathcal{R}}^{(2k)}(U)}_{2}\ge \frac{\norm{\hat{\mathcal{R}}^{(2k)}(U)}_{1}}{d^{k}}\ge\frac{\left|\tr(\hat{\mathcal{R}}^{(2k)}(U))\right|}{d^k}
\ee
where we used the property $|\tr(A)|\le \norm{A}_1$, which can be derived from\cite{watrous2018theory} $|\tr(AB)|\le \norm{A}_{p}\norm{B}_{q}$ where $p^{-1}+q^{-1}=1$, setting $B\equiv \bbbone$, $p=1$ and $q=\infty$. Finally, we obtain:
\be
\mathcal{F}_{\mathcal{E}_H}^{(k)}\ge \frac{\left|\tr(\hat{\mathcal{R}}^{(2k)}(U))\right|^2}{d^{2k}}
\ee
Just recalling that $\tr(\hat{\mathcal{R}}^{(2k)}(U))=|\tr(U)|^{2k}$ we get the desidered bound $(3)$.
\subsection{Definition of generic spectrum}\label{subsecgeneric}
Given $\{E_{n}\}_{n=1}^{d}$ spectrum of a Hamiltonian $H$ on $\mathcal{H}\simeq \mathbb{C}^{d}$; it is said to be {\em generic} iff for any $d\ge l\ge 1$:
\be
\sum_{m=i}^{l}E_{n_{i}}-\sum_{j=1}^{l}E_{m_{j}}\neq 0
\ee
unless $E_{n_{i}}=E_{m_{j}}, \forall i,j=1,\dots, l$ and for some permutation of the indices $n_{i},m_{j}$.
\subsection{Proof of Proposition \ref{prop3}}\label{framepot3}
We need to compute the infinite time average of $|\tr(U)|^{4k}$. Let us write the unitary with its spectral decomposition: $U=\sum_{k}e^{iE_{k}t}\Pi_k$, assuming $\{E_{i}\}_{i=1}^{d}$ be a generic spectrum:
\be
\left|\tr(U)\right|^{4k}=\sum_{\substack{m_1\dots m_{2k}\\n_1\dots n_{2k}}}\exp\left\{it\sum_{i=1}^{2k}E_{m_i}-it\sum_{j=1}^{2k}E_{n_j}\right\}
\ee
Taking the infinite time average the result is zero unless $E_{m_i}=E_{n_j}$ for all $i,j$:
\ba
\hspace{-0.6cm}\overline{\left|\tr(U)\right|^{4k}}^{T}\hspace{-0.5cm}&=&\hspace{-0.3cm}\sum_{\text{pairs}}\sum_{m_1\dots m_{2k}}\sum_{n_1\dots n_{2k}}\delta_{i_1j_1}\cdots \delta_{i_{2k}j_{2k}}+\text{error}\nonumber\\\hspace{-0.3cm}&=&\hspace{-0.3cm}\sum_{\text{pairs}}d^{2k}+\text{error}=(2k)!d^{2k}+O(d^{2k-1})\nonumber\\
\ea
the error come from the fact that we overcount the pairs, e.g. the case $\delta_{i_1j_1}\delta_{i_2j_2}$ overlaps with with $\delta_{i_{1}j_{2}}\delta_{i_2j_1}$ because we are considering twice the case $\delta_{i_1j_1}\delta_{i_{1}j_{2}}\delta_{i_2j_{1}}\delta_{i_2j_2}$; thus the error is $O(d^{2k-1})$. After these considerations we get:
\be
\frac{\overline{\left|\tr(U)\right|^{4k}}^{T}}{d^{2k}}=(2k)!+O(d^{-1})
\ee

\subsection{Proof of  Proposition \ref{prop4}\label{framepot4}}
The $k$-th frame potential of the ensemble $\mathcal{E}_H$ is lower bounded by $|\tr(U)|^{4k}/d^{2k}$, recall $(3)$. Now we should prove that:
\be
\overline{\frac{|\tr(U)|^{4k}}{d^{2k}}}^{\text{GDE}}\ge (2k)!+O(d^{-1})
\ee
So:
\be
|\tr(U)|^{4k}=\sum_{\substack{m_1\dots m_{2k}\\n_1\dots n_{2k}}}\exp\left\{it\sum_{i=1}^{2k}E_{m_i}-it\sum_{j=1}^{2k}E_{n_j}\right\}
\ee
Let's exclude from this sum all the terms such that $E_{m_i}=E_{n_j}$ for any pairs:
\ba
\hspace{-0.6cm}\frac{|\tr(U)|^{4k}}{d^{2k}}\hspace{-0.3cm}&=&\hspace{-0.3cm}(2k)!\nonumber\\\hspace{-0.3cm}&+&\hspace{-0.3cm}d^{-2k}
\hspace{-0.8cm}\sum_{\substack{m_{1}\neq n_1,\dots, n_{2k} \\\dots\dots\dots\dots\dots\dots\dots\\ m_{2k}\neq n_{1},\dots,n_{2k}}}\hspace{-0.7cm}\exp\left\{it\sum_{i=1}^{2k}E_{m_i}-it\sum_{j=1}^{2k}E_{n_j}\right\}\nonumber
\\\hspace{-0.3cm}&+&\hspace{-0.3cm}O(d^{-1})
\ea
see \ref{framepot3} for a discussion regarding the error $O(d^{-1})$. After the ensemble average over GDE\cite{Oliviero2020random}, the second term of the above equation returns a sum of Fourier transforms of Gaussians weighted by positive coefficients depending on the dimension $d$. Hence:
\be
\frac{|\tr(U)|^{4k}}{d^{2k}}\ge (2k)!+O(d^{-1})
\ee
This concludes the proof.

%%%%%%%%%%%%%%%%%%%%%%OTOCLE%%%%%%%%%%%%%%%%%%%%%%%%%%%%%%%%%%%%%
\section{Loschmidt-Echo\label{otocle}} 

\subsection{Proof of Proposition \ref{prop6}}\label{le10}
We first show that, when the perturbation of $H$ leaves the spectrum unchanged, $\text{Sp}(H)=\text{Sp}(H+\delta H)$, then $\mathcal L(t)$ can be viewed as a 2 point auto-correlation function. Any perturbation which leaves the spectrum unchanged can be viewed as a perturbation obtained by rotating the Hamiltonian by a unitary operator close to the identity, say $A\in \mathcal{U}(\mathcal{H})$, $H+\delta H= A^{\dag}HA$; one obtains:
\ba
\mathcal{L}(t)&=&d^{-2}|\tr(e^{iHt}e^{-iA^\dag H At})|^2\nonumber\\&=&d^{-2}|\tr(e^{iHt}A^{\dag}e^{-iHt}A)|^2
\ea
in the last equality we have used the unitarity of $A$ and the series of $e^{iHt}=\sum_{n}\frac{(it)^{n}}{n!}H^{n}$. Twirling the unitary evolution with $G$, $U_{G}=G^{\dag}\exp\{-iHt\}G$ and using $\tr(A)\tr(B)=\tr(A\otimes B)$, one can easily express:
\be
\mathcal{L}(t)=d^{-2}\tr(U_{G}^{\otimes 2}(A\otimes A^{\dag}) U_{G}^{\dag\otimes 2}(A^{\dag}\otimes A))
\ee
It's straightforward to verify $\tr(T_{(13)(24)}A^{\otimes 2}\otimes B^{\otimes 2})=\tr(A^{\otimes 2}B^{\otimes 2})$ and express
\be
 \mathcal{L}(t)=\tr(T_{(13)(24)} (U^{\otimes 2}_{G}\otimes U^{\dag\otimes 2}_{G})A\otimes A^{\dag}\otimes A^{\dag}\otimes A)
 \ee
Inserting $T_{(13)(24)}^{2}\equiv\bbbone$ one gets:
\be
\mathcal{L}(t)=\tr(T_{(13)(24)}(A^{\dag}\otimes A\otimes A\otimes A^{\dag}) (U^{\otimes 2}_{G}\otimes U^{\dag\otimes 2}_{G}))
\ee 
Inserting $T_{(12)}^{2}\equiv \bbbone$ and noting that $[T_{(12)},(U^{\otimes 2}_{G}\otimes U^{\dag\otimes 2}_{G})]=0$ we finally get $(10)$.

%%%%%%%%%%%%%%%%%%%%%%%%%%%%%%%%%%%%%%%%%%%%%%%%%%%%%%%%%%%%%%%%%%%%%%%%%%%%
%ENTANGLEMENT
%%%%%%%%%%%%%%%%%%%%%%%%%%%%%%%%%%%%%%%%%%%%%%%%%%%%%%%%%%%%%%%%%%%%%%%%%%%%

\section{Entanglement}
\subsection{Proof of Proposition \ref{prop7}}\label{subenta}
Starting from the definition of $2-$R\'enyi entropy  ,through the identity $\tr_{A}(\mathcal{O}_{A}^2)=\tr(\mathcal{O}_{A}^{\otimes 2} T_{(A)})$, where $T_{(A)}\equiv T_{A}\otimes \bbbone_{B}^{\otimes 2}$ and $T_{A}$ is the swap operator, it is possible to express the $2-$R\'enyi entropy as:
\be
S_2 =-\log \tr\left[ T_{(A)} U^{\otimes 2} \psi^{\otimes 2}U^{\dag\otimes 2}\right]
\ee 
Again we average over the isospectral unitary evolutions by $U\mapsto G^\dagger U G$ and obtain
\ba
\aver{S_{2}}_{ G} \hspace{-0.3cm}&=&\hspace{-0.3cm}  \langle -\log \tr\left[ T_{(A)} U^{\otimes 2} \psi^{\otimes 2}U^{\dag\otimes 2}\right]\rangle_G
\nonumber\\\hspace{-0.3cm}&\ge&\hspace{-0.3cm}-\log  \tr\left[ T_{(A)} \langle U^{\otimes 2} \psi^{\otimes 2}U^{\dag\otimes 2}\rangle_G\right]\hspace{0.5cm}
\ea
where the lower bound follows by the Jensen inequality by the concavity of the function $-\log$. Now, calling $a\equiv U^{\otimes 2}\psi^{\otimes 2}, b\equiv U^{\dag\otimes 2}T_{(A)}$, and using $\tr[ab]=\tr[T_{(13)(24)}a\otimes b]$, we obtain
\be\label{s2r2}
\aver{S_{2}}_{ G} \ge -\log\tr\left[T_{(13)(24)}\hat{\mathcal R}^{(4)}(U)\psi^{\otimes 2} \otimes T_{(A)}\right]
\ee

\section{Tripartite mutual Information\label{tripartitemutual}}

\subsection{\label{Choi}Choi state: definition and properties}
Let $U\in \mathcal{U}(\mathcal{H})$ be a unitary operator, which decomposed in a basis $\{\ket{i}\}$ reads
$
U=\sum_{i,j}u_{ij}\ket{i}\bra{j}
$.
The Choi isomorphism maps an operator $\mathcal{O}$ into a state $\ket{\mathcal{O}}\in\mathcal{B}(\mathcal{H}^{\otimes 2})$.
The two copies $\mathcal{H}^{\otimes 2}$ of the Hilbert space can be thought of as one lying in the past (input), the other in the future (output)\cite{hosur2016chaos}.
The normalized state corresponding to $U$ reads:
\be 
\ket{U}=d^{-1/2}\sum_{ij} u_{ji} \ket{i}\otimes \ket{j}
\ee 
note that defined $\ket{I}$ as the bell state between the two copies of $\mathcal{H}$:
$
\ket{I}=\frac{1}{\sqrt{d}}\sum_{i} \ket{i}\otimes \ket{i}\
$ one has $
\ket{U}=(\bbbone\otimes U)\ket{I}
$.
The density matrix associated with the Choi state of $U$:
\be
\rho_U=\ket{U}\bra{U}=(\bbbone\otimes U)\ket{I}\bra{I}(\bbbone{}\otimes U^{\dag})
\ee
One important property is that if one traces out the input (output), the resulting state is always maximally mixed:  this reflects the idea that the input and the output are always maximally entangled. Here we can prove a slightly stronger statement.

\begin{prop} Let $f$ be a trace preserving, \textit{unital} $CP-$map; the Choi state $\rho_f=\bbbone\otimes f(\ket{I}\bra{I})$ is such that: 
\be
\tr_{1(2)}(\rho_f)\propto \bbbone_{2(1)}
\ee
where the subscript ${1(2)}$ indicates the first (second) copy of $\mathcal{H}$.
\end{prop}
{\em Proof.}
Let us first prove the statement for $1$. Writing $\rho_{f}$ explicitly and tracing out the input we get
$\tr_{1}(\rho_{f})=d^{-1}\sum_{ij}\tr(\ket{i}\bra{j}_{1})\otimes f(\ket{i}\bra{j}_2)$ from which
one gets $\tr_{1}(\rho_f)\propto f(\bbbone)$, which is the identity since $f$ is unital. To prove the statement for $2$ we need that $f$ is trace preserving:
$\tr_{2}(\rho_f)=d^{-1}\sum_{ij}\ket{i}\bra{j}_{1}\otimes \tr(f(\ket{i}\bra{j}_{2}))=d^{-1}{\bbbone}$, which concludes the proof.
This proposition is important for our purposes: it ensures that taking the average over an ensemble of unitaries, $\mathcal{E}_{H}$ in particular, preserves the properties of the Choi state.

\subsection{$2-$R\'enyi TMI}\label{TMI}
Consider a unitary time evolution $U_{AB\rightarrow CD}$, where $A,B$ and $C,D$ are fixed bipartitions of dimensions $d_A,\dots, d_D$ of past and future time slices of the quantum system: $(\mathcal{H}_{A}\otimes\mathcal{H}_{B})\otimes (\mathcal{H}_C\otimes\mathcal{H}_D)$. The TMI defined through $2-$R\'enyi entropy reads\cite{hosur2016chaos,ding2016conditional}:
\be 
I_{3_{(2)}}=S_2(C)+S_2(D)-S_2(AC)-S_2(AD)
\ee
from the hierarchy of R\'enyi entropy follows $I_{3}\le I_{3_{(2)}}$, where $I_{3}$ is the TMI calculated with the Von Neumann entropy. For the Choi state of $U_{AB\rightarrow CD}$, $\rho_U$, we have:
\be
I_{3_{(2)}}=\log d +\log\tr(\rho_{AC}^{2})+\log\tr(\rho_{AD}^{2})
\label{ref1bis}
\ee
where $\rho_{AC(AD)}=\tr_{BD(BC)}(\rho_U)$.
It is straightforward to see that $I_{3_{(2)}}$ has the following bounds:
\be
-2\log d_{A}\le I_{3_{(2)}}\le 0
\ee
the lower bound is achieved when the information has scrambled, while the upper bound is achieved in the opposite case.

\begin{prop}
The unitaries of the type $\tilde{U}=U_{C}\otimes U_{D}$, where $U_C\in\mathcal{U}(\mathcal{H}_C),\,U_D\in\mathcal{U}(\mathcal{H}_D)$, satisfy:
\be
I_{3_{(2)}}(\tilde{U})=0
\ee
therefore, according to this measure of scrambling, do not scramble the information. 
\end{prop}
{\em Proof.}
First rewrite Eq. $\eqref{ref1bis}$ as
\be 
I_{3_{(2)}}=\log d +\log \tr\left( \rho^{\otimes 2} T_{(A)}T_{(C)}\right)+\log \tr \left(\rho^{\otimes 2} T_{(A)}T_{(D)}\right)
\label{ref2}
\ee 
where $T_{(A)}\equiv T_{A}\otimes \bbbone_{BCD}^{\otimes 2}$ and $T_{A}$ the swap operator on $\mathcal{H}_{A}$, similarly for $T_{(C)}$ and $T_{(D)}$. The Choi state of $\tilde{U}$:
\be
\rho_{\tilde{U}}=(\bbbone_{AB}\otimes U_{C}\otimes U_{D})\ket{I}\bra{I}(\bbbone_{AB}\otimes U_{C}\otimes U_{D})^{\dag}
\ee 
Inserting it in Eq. $\eqref{ref2}$, making use of the cyclic property of the trace and exploting the fact that $[U_{C(D)},T_{D(C)}]=0$, one gets: 
\ba
\hspace{-0.6cm}I_{3_{(2)}}(\tilde{U})\hspace{-0.3cm}&=&\hspace{-0.3cm}\log d +\log \tr\left(\ket{I}\bra{I}^{\otimes 2}T_{(A)}U_{C}^{\dagger\otimes 2}T_{(C)}U_{C}^{\otimes 2}\right)\nonumber\\\hspace{-0.3cm}&+&\hspace{-0.3cm}
\log \tr \left(\ket{I}\bra{I}^{\otimes 2} T_{(A)}U_{D}^{\dagger\otimes 2} T_{(D)}U_{D}^{\otimes 2}\right)
\ea
since $\left[U_{C}^{\otimes 2},P_{C}\right]=0$, and $\left[U_{D}^{\otimes 2},P_{D}\right]=0$, we obtain 
\ba
\hspace{-0.6cm}I_{3_{(2)}}(\tilde{U})\hspace{-0.3cm}&=&\hspace{-0.3cm}\log d +\log \tr\left( \ket{I}\bra{I}^{\otimes 2} T_{(A)}T_{(C)}\right)
\nonumber\\\hspace{-0.3cm}&+&\hspace{-0.3cm}\log \tr\left( \ket{I}\bra{I}^{\otimes 2} T_{(A)}T_{(D)}\right)
\ea
at this point a straightforward calculation shows that $I_{3_{(2)}}(\tilde{U})=0$.
\subsection{Proof of Eq. $\eqref{tracedTMI}$ and Proposition $\eqref{prop8}$\label{prooftracedTMI}}
We start from Eq. \eqref{ref1bis} and rewrite it as Eq. \eqref{ref2}.
where $\rho_{U}=(\bbbone{}_{AB}\otimes U_{CD})\ket{I}\bra{I}(\bbbone{}_{AB}\otimes U^{\dag}_{CD})$. Let us set $A=C$ and $B=D$.
We need two facts. First, note that in this settings $\ket{I}\bra{I}=\psi_{AC}\otimes \psi_{BD}$ where $\psi_{AC},\psi_{BD}$ are Bell states. Second, let $T$ be the swap operator and let $\psi\in\mathcal{H}$ a pure state, then
$
T\psi^{\otimes 2}=\psi^{\otimes2}
$.
Now let us focus on the second term of Eq. $\eqref{ref2}$; using the above identity we have:
\be
T_{(A)}T_{(C)}\psi_{AC}^{\otimes2}=\psi_{AC}^{\otimes2}\implies T_{(A)}\psi_{AC}^{\otimes2}=T_{(C)}\psi_{AC}^{\otimes2}
\ee
in this way we can trace out  $\mathcal{H}_A\otimes\mathcal{H}_B$ and obtain that:
\be
\log \tr \left(\rho_{U}^{\otimes 2} T_{(A)}T_{(C)}\right)={d^{-2}}\log \tr \left( U_{CD}^{\otimes 2}T_{(C)} U_{CD}^{\dag\otimes 2} T_{(C)}\right)
\ee
where the factor $d^{-2}$ comes from having traced out the Bell states $\tr_{A(B)}(\psi_{AC(BD)}^{\otimes 2})=d_{C(D)}^{-2}\bbbone_{C(D)}$. For the second term Eq. \eqref{ref2} we can play the same game and obtain:
\be
\log \tr \left(\rho_{U}^{\otimes 2} T_{(A)}T_{(D)}\right)={d^{-2}}\log \tr \left( U_{CD}^{\otimes 2}T_{(C)} U_{CD}^{\dag\otimes 2} T_{(D)}\right)
\ee
to prove Eq. $(13)$ we use $\tr(ab)=\tr(T_{(13)(24)}a\otimes b)$ if $a,b\in\mathcal{H}^{\otimes 2}$, then we take the isospectral twirling over $U_{G}=G^{\dag}\exp\{-iHt\}G$ and use the Jensen inequality to upper bound $I_{3_{(2)}}$.

\end{document}